**Spin–orbit-enabled Fermi-surface splitting in noncollinear antiferromagnetic SmBi**

Long Zhang[1†], Ming Cheng[2*], Jingyu Li[3*], Honghao Wan[4,5*], Xiaoyuan Zhou[4], Mingquan He[1,4], Aifeng Wang[1,4], Yuping Sun[2,6,7], Dong-Hui Xu[4,5*], Huixia Fu[4,5*], Youguo Shi[3*], Xuan Luo[2*], Yisheng Chai[1,4†]

[1]*Low Temperature Physics Laboratory, College of Physics, Chongqing University, Chongqing 401331, China.*

[2]*Key Laboratory of Materials Physics, Institute of Solid State Physics, HFIPS, Chinese Academy of Sciences, Hefei 230031, China.*

[3]*Beijing National Laboratory for Condensed Matter Physics, Institute of Physics, Chinese Academy of Sciences, Beijing 100190, China.*

[4]Center *of Quantum Materials and Devices, Chongqing University, Chongqing 401331, China.*

[5] *Department of Physics and Chongqing Key Laboratory for Strongly Coupled Physics, College of Physics, Chongqing University, Chongqing 401331, China.*

[6]*Anhui Key Laboratory of Low-Energy Quantum Materials and Devices, High Magnetic Field Laboratory, HFIPS, Chinese Academy of Sciences, Hefei 230031, China.*

[7]*Collaborative Innovation Center of Advanced Microstructures, Nanjing University, Nanjing 210093, People's Republic of China.*

[*]donghuixu@cqu.edu.cn, [*]hxfu@cqu.edu.cn, [*]ygshi@iphy.ac.cn, [*]xluo@issp.ac.cn, [†]yschai@cqu.edu.cn

**Spin-split electronic structures in compensated antiferromagnets are commonly sought in the nonrelativistic limit, where magnetic order lifts spin degeneracy without spin–orbit coupling (SOC). Whether SOC can instead be the indispensable symmetry-breaking ingredient remains largely unexplored. Here we combine quantum oscillations detected by ultrahigh-sensitivity ac magnetostriction, magnetic-symmetry analysis and first-principles calculations to resolve the bulk Fermi-surface evolution of SmBi across two successive antiferromagnetic (AFM) transitions. New oscillation branches emerge below $T_N$ and undergo a further reconstruction below $T^*$, whereas isostructural SmSb shows no comparable change. For the candidate noncollinear orders of SmBi, breaking global parity–time ($\mathcal{PT}$) symmetry is insufficient in the nonrelativistic limit because residual spin-space**

**symmetries protect twofold band degeneracy; conversely, SOC alone cannot lift the degeneracy of the centrosymmetric paramagnetic (PM) phase. Only the coexistence of noncollinear order and SOC locks spin to the lattice and removes the residual protection. SmBi therefore realizes a cooperative, relativistic route to spin-split Fermi surfaces, broadening unconventional magnetism beyond systems whose splitting is already present in the nonrelativistic limit.**

Spin-split electronic structures in compensated antiferromagnets have become a central target of unconventional magnetism. Antiferromagnets combine negligible stray fields with ultrafast dynamics and robustness against magnetic perturbations[1]; lifting their band degeneracy can additionally produce spin-polarized carriers and transport responses without a net moment. In conventional collinear antiferromagnets, symmetries such as $\mathcal{PT}$ or time reversal followed by a translation ($\mathcal{T}\tau$) can protect Kramers-like band degeneracies. Spin-group symmetry has revealed routes around this constraint, most prominently the nonrelativistic, exchange-driven splitting of collinear altermagnets[2,3]. Experiments have since resolved altermagnetic splitting in MnTe[4], plaid-like splitting generated by noncoplanar order in $MnTe_2$[5], a metallic d-wave altermagnet in $KV_2Se_2O$[6], and odd-parity p-wave splitting associated with a commensurate spin helix[7]. These advances establish spin-split compensated magnets as a broad symmetry-defined landscape rather than a single magnetic phase.

Much of this progress has deliberately focused on spin splitting that survives when SOC is removed, because the nonrelativistic limit isolates magnetic exchange and spin symmetry from conventional relativistic effects. Although SOC can further lift nodal degeneracies in altermagnets[4], their defining momentum-dependent splitting is already present without SOC. This leaves open a complementary possibility: SOC may be the indispensable symmetry-breaking ingredient rather than a perturbation to an already spin-split state. Breaking global $\mathcal{PT}$ symmetry by noncollinear order does not by itself guarantee spin splitting. When spin and lattice rotations are decoupled, residual spin-space operations can enforce twofold degeneracy throughout the Brillouin zone. SOC reduces this enlarged symmetry by locking spin to the lattice[8], yet SOC alone cannot split the bands of a centrosymmetric paramagnet in which $\mathcal{P}$ and $\mathcal{T}$ remain intact. A compensated magnet can therefore develop spin splitting cooperatively only when noncollinear order and SOC coexist, even though neither ingredient is sufficient alone. Bulk experimental evidence for this cooperative relativistic mechanism remains scarce.

Quantum oscillations (QOs) provide a stringent bulk test of the resulting electronic reconstruction: through the Onsager relation, their frequencies measure extremal Fermi-surface (FS) areas with high precision. If a symmetry-breaking transition lifts a band degeneracy, paired or newly resolved extremal orbits may emerge. QOs alone, however, do not determine spin character; a compelling assignment requires their combination with magnetic-symmetry analysis and electronic-structure calculations. Recent QO studies have tested the proposed altermagnetic bulk state of $RuO_2$[9] and reported spin-split AFM FSs in CrSb[10–12]. The outstanding question is whether the same bulk approach can expose an SOC-enabled reconstruction in a compensated noncollinear antiferromagnet whose bands remain fully degenerate in the nonrelativistic limit.

Rare-earth monopnictides offer a clean setting for this test. Their rock-salt lattice combines simple semimetallic Fermi surfaces, strong atomic SOC and long-range AFM order, while high-quality crystals support well-resolved QOs[13]. SmSb and SmBi are particularly instructive: they are isostructural and electronically similar in their PM states[14], yet their ordered phases differ sharply. SmSb undergoes a single AFM transition near 2.4 K and exhibits anomalous QO amplitudes[15] but no resolvable FS reconstruction across the transition[16], consistent with a degeneracy-preserving collinear state. SmBi, by contrast, undergoes two successive AFM transitions near 9 and 7 K and develops additional low-temperature QO branches[17]. Whether these branches are bulk, how they evolve through the two phases and which symmetry mechanism produces them have remained unresolved.

Here we use ultrahigh-sensitivity ac magnetostriction based on a composite magnetoelectric technique to track the magnetic transitions and QOs of SmBi within the same bulk thermodynamic channel[18–22]. New oscillation branches appear below $T_N$, and the spectrum reconstructs again below $T^*$, revealing sequential evolution of the bulk FS. Magnetic-symmetry analysis identifies a higher-symmetry cubic Kouvel–Kasper (KK) order (Type-I) and a lower-symmetry Type-III order as candidates for the two phases[23–25], and first-principles calculations yield spin-split Type-I bands only when SOC is included. Crucially, both candidate orders break global $\mathcal{PT}$ symmetry yet retain spin-space symmetries that enforce twofold band degeneracy throughout the Brillouin zone in the SOC-free limit; conversely, the PM state remains doubly degenerate with SOC because $\mathcal{P}$ and $\mathcal{T}$ are both present. The coexistence of noncollinear order and SOC therefore provides a symmetry-based account of the observed reconstruction. The absence of a comparable reconstruction in isostructural SmSb provides a control against antiferromagnetism alone. Our results identify SmBi as a bulk platform for cooperative SOC-enabled spin splitting and broaden the search for spin-split compensated magnets beyond systems whose splitting is already present in the nonrelativistic limit.

Figure 1a shows the NaCl-type crystal structure of SmBi and the cubic KK-type noncollinear magnetic configuration proposed for the AFM state[25]. Because natural Sm strongly absorbs neutrons, direct determination of the magnetic structure by neutron diffraction is challenging. We therefore begin with the macroscopic magnetic properties. The temperature-dependent susceptibility exhibits two successive transitions at $T_N \approx 9$ K and $T^* \approx 7$ K, defining AFM-I ($T^* < T < T_N$) and AFM-II ($T < T^*$) (Fig. 1b). The isothermal magnetization $M(H)$ remains linear at all measured temperatures up to 9 T (Fig. 1c), consistent with compensated magnetic order in both phases. Quantum oscillations (QOs) become visible above approximately 2.5 T. Fast Fourier transform (FFT) analysis (Extended Data Fig. 1) resolves three frequencies below 9 K: $\delta \approx 40$ T, $\varepsilon \approx 102$ T and $\alpha \approx 327$ T, in agreement with previous work[17]. At 9 K, only a weak frequency near 687 T is resolved; comparison with density functional theory (DFT) assigns it to the $\beta$ hole pocket[17]. The marked change in the QO spectrum below $T_N$ points to a reconstruction of the bulk Fermi surface (FS) and motivates consideration of noncollinear magnetic orders.

To resolve the magnetic and FS evolution in the two ordered phases more sensitively, we measure the ac magnetostrictive coefficient $(\mathrm{d}\lambda/\mathrm{d}H)_{ac}$ using the composite magnetoelectric (ME) configuration in Fig. 2a. Figures 2b and 2c show the real part $\mathrm{d}\lambda'/\mathrm{d}H$ and the imaginary part $\mathrm{d}\lambda''/\mathrm{d}H$, respectively, as functions of temperature for $H \| [001]$. Four features stand out. First, $\mathrm{d}\lambda'/\mathrm{d}H$ increases step-like from nearly zero on cooling from the PM state into AFM-I, as expected at the onset of AFM order[18]. A pronounced kink at $T^*$ marks the transition into AFM-II. The absence of the peak-like anomaly characteristic of a PM–ferromagnetic transition in this technique argues against ferromagnetic order. Second, $\mathrm{d}\lambda''/\mathrm{d}H$ differs sharply between the two AFM phases: it is nearly zero in AFM-I but develops a broad negative minimum in AFM-II, with the largest magnitude near 4 T and gradual suppression above 10 T. We associate this dissipative response with the multidomain state expected from degenerate chiral configurations in AFM-II, as discussed below. Third, pronounced oscillations emerge in $\mathrm{d}\lambda'/\mathrm{d}H$ above approximately 5 T for $T < T_N$, with frequencies that evolve with temperature[26]. Finally, neither transition shifts detectably up to 13.8 T, indicating that both ordered phases are robust over the field range studied.

We next analyse the QOs across the PM, AFM-I and AFM-II regimes using field-dependent $(\mathrm{d}\lambda/\mathrm{d}H)_{ac}$. We do not use the oscillations alone to assign a magnetic symmetry; instead, we treat the evolution of the QO branches as bulk evidence for FS reconstruction and interpret it together with symmetry analysis and DFT. Figure 3a shows $\mathrm{d}\lambda'/\mathrm{d}H$ between 2 and 20 K. Both the oscillatory pattern and the nonoscillatory magnetic background change across the three regimes. After subtraction of a smooth background, the oscillatory component $\Delta(\mathrm{d}\lambda'/\mathrm{d}H)$ remains detectable up to 20 K, well above $T_N$, whereas field-dependent magnetization resolves only much weaker oscillations up to 9 K (Extended Data Fig. 1).

The FFT contour plot in Fig. 3c summarizes the temperature evolution of the QO spectrum. In the PM phase, two nearly temperature-independent frequencies at 376 and 687 T are resolved and assigned to the $\alpha$ electron pocket and $\beta$ hole pocket, respectively. On entering AFM-I, both branches remain continuous across $T_N$. A weak branch just above $\alpha$, denoted $\alpha'$ (Extended Data Fig. 2a), and a pronounced low-frequency branch $\delta$ below 50 T nevertheless emerge and shift with temperature. In AFM-II, $\alpha$, $\beta$ and $\delta$ evolve continuously through $T^*$, whereas $\alpha'$ resolves into two branches, $\mu$ and $\nu$, whose frequencies shift in opposite directions on cooling (Fig. 3c). An additional $\varepsilon$ branch appears just above 100 T below approximately 4 K. Because the composite ME response is a bulk thermodynamic quantity, the $\delta$ and $\varepsilon$ signals are unlikely to originate from the surface states proposed in the earlier de Haas–van Alphen study[17]. They were also not resolved by angle-resolved photoemission spectroscopy at 6 K, plausibly because the associated momentum-space changes are below the experimental resolution[27,28]. Weak high-frequency features assigned to $\alpha_2$, $\alpha_3$ and $\gamma$ are observed above 1 kT (Extended Data Fig. 2b). Taken together, the data establish sequential reconstruction of the bulk FS across the two AFM transitions. By contrast, isostructural SmSb, which undergoes only one AFM transition, shows neither additional fundamental branches nor resolvable frequency shifts across its Néel temperature (Supplementary Fig. S3). This control is consistent with a degeneracy-preserving collinear state in SmSb and links the reconstruction in SmBi to its lower-symmetry magnetic order rather than to antiferromagnetism alone.

To identify candidate spin configurations for the two magnetic phases, we examine diagonal-spin noncollinear AFM structures on the fcc Sm sublattice. Crystal-field considerations favour a [111] easy axis for $Sm^{3+}$ in SmBi[24]. Yamamoto and Nagamiya classified the diagonal-spin configurations allowed on the fcc sublattice of the rock-salt structure into Type-I, Type-II and Type-III categories[23] (Fig. 4). Within this framework, Type-I and Type-III are the experimentally relevant candidates. Type-I has propagation vectors $Q$ = (π, 0, 0), (0, π, 0) and (0, 0, π), and corresponds to the cubic KK configuration proposed by Hulliger et al. from PM susceptibility[24]. Type-III has propagation vectors $Q$ = (π, 0, π/2), (0, π/2, π) and (π/2, π, 0), and a larger magnetic unit cell. For Type-I, the magnetic space group (MSG) and magnetic point group (MPG) are Pn-3m′ and m-3m′, respectively. This state is fully compensated and lacks the global degeneracy protection of a simple collinear $\mathcal{PT}$-symmetric AFM, although spin-space symmetries still enforce nonrelativistic degeneracy, as discussed below. For Type-III, the MSG is $P_I4_132$ or its enantiomorphic partner $P_I4_332$, and the MPG is 4321′. Type-III lacks spatial inversion and hence global $\mathcal{PT}$ protection, and it has lower symmetry than Type-I. Consistent with Landau theory, in which a secondary transition generally lowers the symmetry, we assign the higher-symmetry Type-I state to AFM-I and the lower-symmetry Type-III state to AFM-II. Because Type-III breaks inversion, it admits two chiral domains associated with the enantiomorphic pair $P_I4_132$ and $P_I4_332$. Their coexistence offers a natural explanation for the negative $d\lambda''/dH$ response observed only in AFM-II. The opposite signs of the real and imaginary components of $(d\lambda/dH)_{ac}$ are reminiscent of magnetoelastic responses in phase-coexistence regimes, including first-order magnetic transitions[18,19] and the intermediate state of type-I superconductors[29].

To determine how the candidate magnetic structures affect band degeneracy, we performed DFT calculations for the PM state, a representative collinear AFM state and the Type-I noncollinear state. Direct calculations for Type-II and Type-III are substantially more demanding because of their larger magnetic unit cells; we therefore assess these phases primarily through magnetic-space-group and spin-space-group analyses. All calculated bands were unfolded into the primitive-cell Brillouin zone. In the PM state, DFT including SOC yields doubly degenerate bands throughout the Brillouin zone because inversion and time-reversal symmetries are both present (Supplementary Fig. S5). Three bands cross the Fermi energy $E_F$, producing one electron pocket at X and two hole pockets around Γ, identified with $\alpha$, $\beta$ and $\gamma$ (Extended Data Fig. 3a). Experimentally, $\alpha$ and $\beta$ are resolved in $d\lambda'/dH$ in the PM phase, whereas $\gamma$ is not, probably because the accessible PM temperature range is limited by the relatively high $T_N$ and the $\gamma$ orbit has a larger effective mass (Supplementary Fig. S4). A representative collinear AFM state preserves combined $\mathcal{PT}$ symmetry and remains doubly degenerate (Supplementary Fig. S5), consistent with the unchanged QO frequencies of SmSb across $T_N$ (Supplementary Fig. S3). By contrast, the Type-I noncollinear calculation including SOC exhibits split bands (Supplementary Fig. S5), providing a microscopic realization of the splitting inferred for AFM-I. This mechanism differs from nonrelativistic altermagnetism. Spin-space-group analysis shows that Type-I retains spin-space operations including $\{C_{2x} \parallel E \mid 0\ \tfrac{1}{2}\ \tfrac{1}{2}\}$; their symmetry algebra enforces twofold band degeneracy and vanishing spin polarization $S(k) = 0$ when SOC is absent[30]. SOC reduces this enlarged symmetry by locking spin to the lattice; because Type-I already breaks global $\mathcal{PT}$, the residual degeneracy can then be lifted. Type-III is analogous: a $\mathcal{T}\tau$-related spin-space symmetry enforces $S(k) = 0$ and double degeneracy in the SOC-free limit, whereas SOC permits splitting once global $\mathcal{PT}$ is absent. Thus, spin splitting in both candidate phases is cooperative and SOC-enabled rather than nonrelativistic. Because the itinerant states near $E_F$ hybridize only weakly with the localized Sm 4f moments, the resulting splitting is expected to remain modest, consistent with the small FS reconstruction observed experimentally.

With this symmetry picture, we next relate the reconstructed QO branches to candidate FS pockets. The emergence of $\delta$ and $\varepsilon$, together with the appearance of a second $\alpha$-like branch $\alpha'$, constitutes the principal reconstruction below $T_N$. The calculated PM FS and measured QO frequencies agree qualitatively. A quantitative comparison in AFM-I is limited by the difficulty of treating the Sm 4f states accurately, but the calculated Type-I band splitting provides a natural origin for the two nearby electron-pocket orbits $\alpha$ and $\alpha'$. Below $T^*$, the further symmetry reduction in Type-III may distort $\alpha'$ from an ellipsoid into a peanut-like pocket, analogous to calculations for ferromagnetic CeBi[31]. Such a pocket has two extremal cross-sections, $\mu$ and $\nu$, which can evolve in opposite directions with temperature (right panel of Extended Data Fig. 3c). This scenario explains why the $\alpha$ family contains three observable frequencies, $\alpha$, $\mu$ and $\nu$, rather than only the two branches expected from a simple splitting.

The small extremal areas associated with $\delta$ and $\varepsilon$ point to an origin distinct from the principal PM pockets. Their temperature-dependent frequency shifts provide a further empirical distinction. The $\alpha$, $\beta$, $\mu$ and $\nu$ branches follow similar $\Delta F \propto T^4$ behaviour, whereas $\delta$ follows $\Delta F \propto T^3$ (Extended Data Figs. 1c and 2d). This contrast suggests that $\alpha$, $\beta$, $\mu$ and $\nu$ respond to a common magnetic reconstruction, whereas $\delta$ arises from a different reconstructed band. The microscopic origin of $\varepsilon$ is less constrained, particularly in the lower-symmetry Type-III phase. Overall, the sequential Type-I-to-Type-III evolution demonstrates how compensated noncollinear order can reshape SOC-split bulk FSs in stages.

Finally, SmBi shows Kondo-like transport near $T_N$[17], so hybridization between the Sm 4f moments and itinerant carriers could contribute to the FS evolution. A conventional coherent Kondo crossover would be expected to produce a broad, largely continuous reconstruction as f-electron spectral weight enters the FS. In SmBi, however, the new branches appear at the magnetic transitions and, with the exception of $\nu$, their frequencies decrease on cooling below Kondo temperature $T_K \approx 9.2$ K rather than showing a uniform expansion. These features do not exclude Kondo hybridization, but they argue against it as the sole origin of the reconstruction and favour magnetic symmetry breaking as the dominant driver.

In summary, ac magnetostriction resolves a sequential bulk FS reconstruction across two AFM transitions in SmBi. The emergence and evolution of the QO branches, interpreted together with symmetry analysis and first-principles calculations, are consistent with the lifting of band degeneracy by the candidate Type-I and Type-III noncollinear orders. In both states, spin-space symmetries protect twofold degeneracy without SOC; only the coexistence of noncollinear order and SOC removes the residual protection. SmBi therefore provides a bulk platform for cooperative, SOC-enabled spin-split fermiology in a compensated antiferromagnet, distinct from the nonrelativistic altermagnetic route. More broadly, these results identify SOC-coupled noncollinear antiferromagnets as a promising materials space for symmetry-engineered spin splitting.

**References**

1. Baltz, V. *et al.* Antiferromagnetic spintronics. *Rev. Mod. Phys.* **90**, 015005 (2018).
2. Šmejkal, L., Sinova, J. & Jungwirth, T. Beyond conventional ferromagnetism and antiferromagnetism: a phase with nonrelativistic spin and crystal rotation symmetry. *Phys. Rev. X* **12**, 031042 (2022).
3. Šmejkal, L., Sinova, J. & Jungwirth, T. Emerging research landscape of altermagnetism. *Phys. Rev. X* **12**, 040501 (2022).
4. Krempaský, J. *et al.* Altermagnetic lifting of Kramers spin degeneracy. *Nature* **626**, 517–522 (2024).
5. Zhu, Y.-P. *et al.* Observation of plaid-like spin splitting in a noncoplanar antiferromagnet. *Nature* **626**, 523–528 (2024).
6. Jiang, B. *et al.* A metallic room-temperature d-wave altermagnet. *Nat. Phys.* **21**, 754–759 (2025).
7. Yamada, R. *et al.* A metallic p-wave magnet with commensurate spin helix. *Nature* **646**, 837–842 (2025).
8. Liu, Y. *et al.* Symmetry classification of magnetic orders using oriented spin space groups. *Nature* **652**, 869–873 (2026).
9. Wu, Z. *et al.* Fermi surface of $RuO_2$ measured by quantum oscillations. *Phys. Rev. X* **15**, 031044 (2025).
10. Long, M. *et al.* 3D bulk-resolved g-wave magnetic order parameter symmetry in the metallic altermagnet CrSb. Preprint at https://arxiv.org/abs/2601.14526 (2026).

11. Terashima, T. *et al.* Altermagnetic spin-split Fermi surfaces in CrSb revealed by quantum oscillation measurements. Preprint at https://arxiv.org/abs/2601.19105 (2026).
12. Naduvile Thadathil, S. *et al.* Fermi-surface studies of altermagnetic CrSb from Shubnikov–de Haas oscillations. Preprint at https://arxiv.org/abs/2602.24033 (2026).
13. Duan, X. *et al.* Tunable electronic structure and topological properties of LnPn (Ln=Ce, Pr, Sm, Gd, Yb; Pn=Sb, Bi). *Commun. Phys.* **1**, 71 (2018).
14. Wu, Z. *et al.* Probing the origin of extreme magnetoresistance in Pr/Sm mono-antimonides/bismuthides. *Phys. Rev. B* **99**, 035158 (2019).
15. Wu, F. *et al.* Anomalous quantum oscillations and evidence for a non-trivial Berry phase in SmSb. *npj Quantum Mater.* **4**, 20 (2019).
16. Zhang, W. *et al.* Tunable non-Lifshitz–Kosevich temperature dependence of Shubnikov–de Haas oscillation amplitudes in SmSb. *npj Quantum Mater.* **8**, 55 (2023).
17. Sakhya, A. P., Paulose, P. L., Thamizhavel, A. & Maiti, K. Evidence of nontrivial Berry phase and Kondo physics in SmBi. *Phys. Rev. Mater.* **5**, 054201 (2021).
18. Zhang, Y. *et al.* Observation of enhanced ferromagnetic spin-spin correlations at a triple point in quasi-two-dimensional magnets. *Phys. Rev. B* **107**, 134417 (2023).
19. Mi, X. *et al.* Precisely tracking critical spin fluctuations using the ac magnetostriction coefficient: a case study on the Kitaev spin liquid candidate $Na_3Co_2SbO_6$. *Phys. Rev. B* **111**, 014417 (2025).
20. Zhang, L. *et al.* Magnetic phase diagram of $Cr_2Te_3$ revisited by ac magnetostrictive coefficient. *Appl. Phys. Lett.* **127**, 012413 (2025).
21. Zhang, L. *et al.* Comprehensive investigation of quantum oscillations in semimetal using an ac composite magnetoelectric technique with ultrahigh sensitivity. *npj Quantum Mater.* **9**, 11 (2024).
22. Lu, Y. *et al.* Fermi surface structure revealed by quantum oscillations in half-Heusler semimetal LuAuSn. *Rare Met.* **45**, e70109 (2026).
23. Yamamoto, Y. & Nagamiya, T. Spin arrangements in magnetic compounds of the rocksalt crystal structure. *J. Phys. Soc. Jpn.* **32**, 1248 (1972).
24. Hulliger, F., Natterer, B. & Rüegg, K. Low-temperature properties of Samarium monopnictides. *Z. Phys. B* **32**, 37 (1978).

25. Kouvel, J. S. & Kasper, J. S. Long-range antiferromagnetism in disordered Fe-Ni-Mn alloys. *J. Phys. Chem. Solids* **24**, 529 (1963).
26. Wang, J.-F. *et al.* Quantum oscillations in the magnetic Weyl semimetal NdAlSi arising from strong Weyl fermion-4f electron exchange interaction. *Phys. Rev. B* **108**, 024423 (2023).
27. Sakhya, A. P. *et al.* Behavior of gapped and ungapped Dirac cones in the antiferromagnetic topological metal SmBi. *Phys. Rev. B* **106**, 085132 (2022).
28. Kushnirenko, Y. *et al.* Rare-earth monopnictides: Family of antiferromagnets hosting magnetic Fermi arcs. *Phys. Rev. B* **106**, 115112 (2022).
29. Yuan, M. *et al.* Collective resonance of superconducting/normal domain walls in the intermediate state of type-I superconductor. Preprint at https://arxiv.org/abs/2604.17333 (2026).
30. Song, Z. *et al.* Constructions and applications of irreducible representations of spin-space groups. *Phys. Rev. B* **111**, 134407 (2025).
31. Huan, S. *et al.* Magnetotransport evidence for the nontrivial topological states in the fully spin-polarized Kondo semimetal CeBi. *J. Alloys Compd.* **875**, 159993 (2021).

## Methods

### Crystal growth

Single crystals of SmBi were grown by a self-flux method.

### Ac magnetostriction measurements

Ac magnetostriction-coefficient measurements were performed using a composite magnetoelectric (ME) technique with the dc magnetic field $H$ applied along [001]. A commercial sample probe (MultiFields Technology), shown schematically in Fig. 2a, was used. A PMN–PT single crystal served as the piezoelectric transducer. A small ac field $H_{ac} \parallel H$ was superimposed on the dc field to drive an ac magnetostrictive response $\lambda_{ac}$. The sample and transducer were bonded with silver epoxy, which transferred strain and provided electrodes for measuring the resulting ac voltage $V_{ME}$ across the PMN–PT. Over the temperature and field ranges studied, $V_{ME}$ is proportional to the ac magnetostrictive coefficient $(\mathrm{d}\lambda/\mathrm{d}H)_{ac}$:

$$V_{\mathrm{ME}} \propto \frac{\lambda_{\mathrm{ac}}}{H_{\mathrm{ac}}} \sim \left(\frac{\mathrm{d}\lambda}{\mathrm{d}H}\right)_{\mathrm{ac}} = \frac{\mathrm{d}\lambda'}{\mathrm{d}H} + i\frac{\mathrm{d}\lambda''}{\mathrm{d}H},$$

Here $d\lambda'/dH$ and $d\lambda''/dH$ denote the real and imaginary components, respectively. Because the proportionality factor depends on the mechanical coupling between sample and transducer, the measured signal is reported in arbitrary units and used as a relative measure of $(d\lambda/dH)_{ac}$.

**Magnetization measurements**

Dc magnetization and magnetic susceptibility were measured using a Quantum Design Physical Property Measurement System equipped with a vibrating-sample magnetometer option.

**FFT analysis**

FFT spectra were calculated using a rectangular window, so that no additional weighting from a tapered window function was applied. Field intervals were selected to include as many oscillation periods in inverse field as possible. Zero-padding was used to obtain a more finely sampled frequency axis; it does not increase the intrinsic frequency resolution.

**Density functional theory calculations**

Electronic band structures were calculated within DFT using the Vienna Ab initio Simulation Package (VASP). A plane-wave cutoff of 300 eV and a Γ-centred 9 × 9 × 9 *k*-point mesh were used, giving total-energy convergence better than 0.01 eV per cell. Exchange and correlation were treated using the Perdew–Burke–Ernzerhof functional within the generalized-gradient approximation, together with projector-augmented-wave potentials for Sm and Bi. In the nonmagnetic calculation used to model the PM itinerant bands, the Sm 4f electrons were treated as part of the frozen core using an 11-valence-electron potential. All AFM calculations used a 16-valence-electron potential that includes the 4f electrons; noncollinear magnetism and SOC were included as required. The Fermi energy of the magnetic calculations was aligned to the itinerant-band filling of the nonmagnetic calculation. VASPKIT was used for post-processing and iFermi for FS visualization. Further computational details are provided in the Supplementary Information.

**Acknowledgements**

This work was supported by the National Natural Science Foundation of China (Grant Nos. 12227806, 12374081, 12274412, 11774399 and 11474330), the National Key Research and Development Program of China (Grant Nos. 2021YFA1600200 to X.L. and Y.P.S., and 2023YFA1607402 to X.L.) and the Beijing National Laboratory for Condensed Matter Physics (Grant No. 2025BNLCMPKF015 to H.F.). Y.S.C. acknowledges support from the Open Research Fund of the Pulsed High Magnetic Field Facility (Grant No. WHMFC2024007), Huazhong University of Science and Technology and the Beijing National Laboratory for Condensed Matter Physics. We thank Y. Liu of the Analytical and Testing Center, Chongqing University, for assistance.

**Author contributions**

L.Z., J.L., M.C. and H.W. contributed equally to this work. L.Z. performed the ac magnetostriction and magnetization measurements, analysed the data and drafted the manuscript. J.L., M.C., Y.G.S. and X.L. grew and characterized the single crystals. H.W., H.F. and D.-H.X. performed the first-principles calculations and symmetry analysis. X.Z. provided experimental facilities. Y.S.C., D.-H.X., H.F., Y.G.S., X.L., Y.P.S., A.W. and M.H. conceived and supervised the project and revised the manuscript. All authors discussed the results and approved the final manuscript.

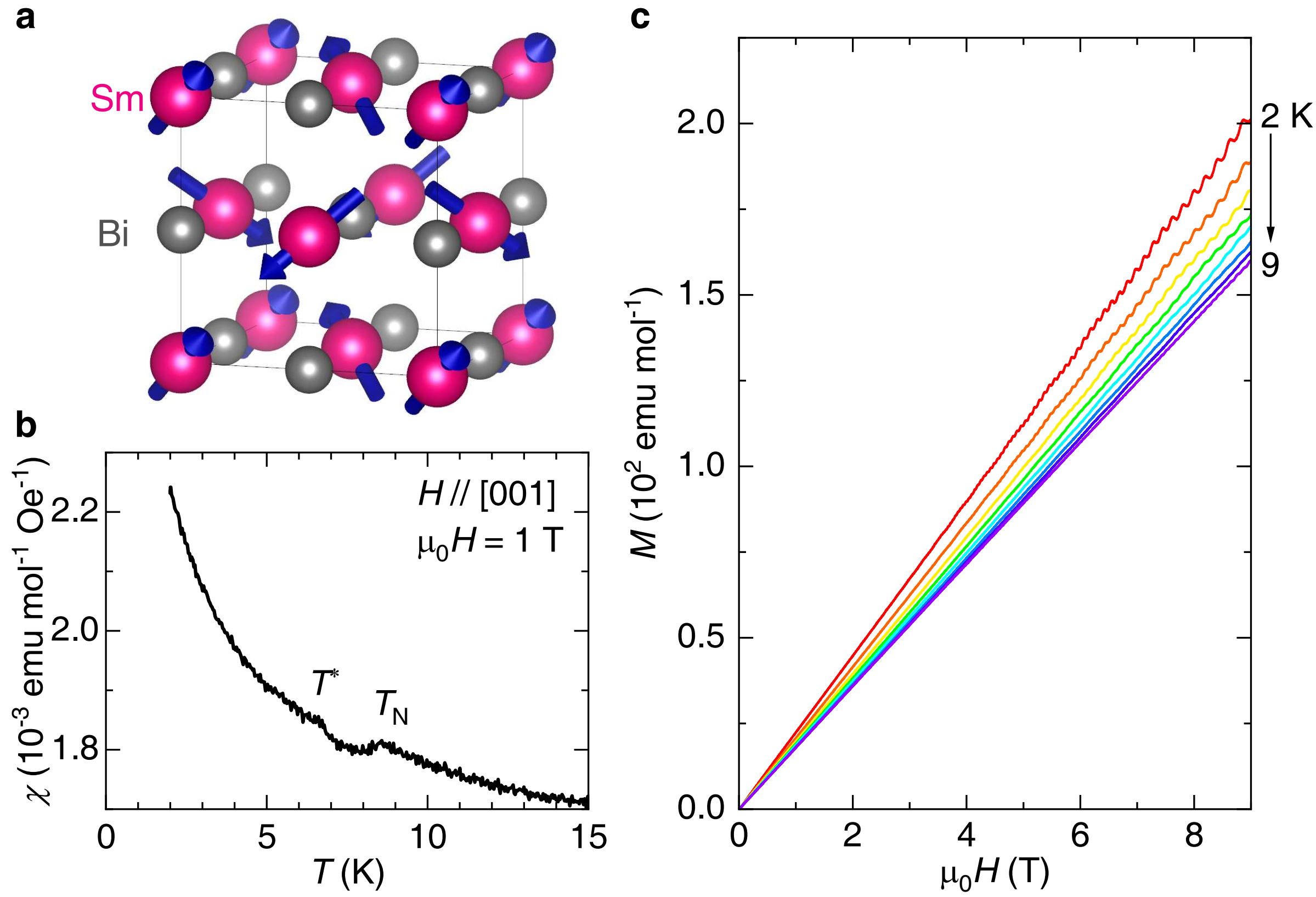

**Fig. 1 | Crystal and magnetic structures and basic magnetic properties of SmBi. a,** NaCl-type crystal structure and proposed cubic KK-type Type-I noncollinear AFM configuration of SmBi (ref. 25), with four threefold rotation axes along the four equivalent [111] directions. **b,** Temperature-dependent magnetic susceptibility for $H \parallel [001]$. **c,** Field-dependent magnetization at selected temperatures.

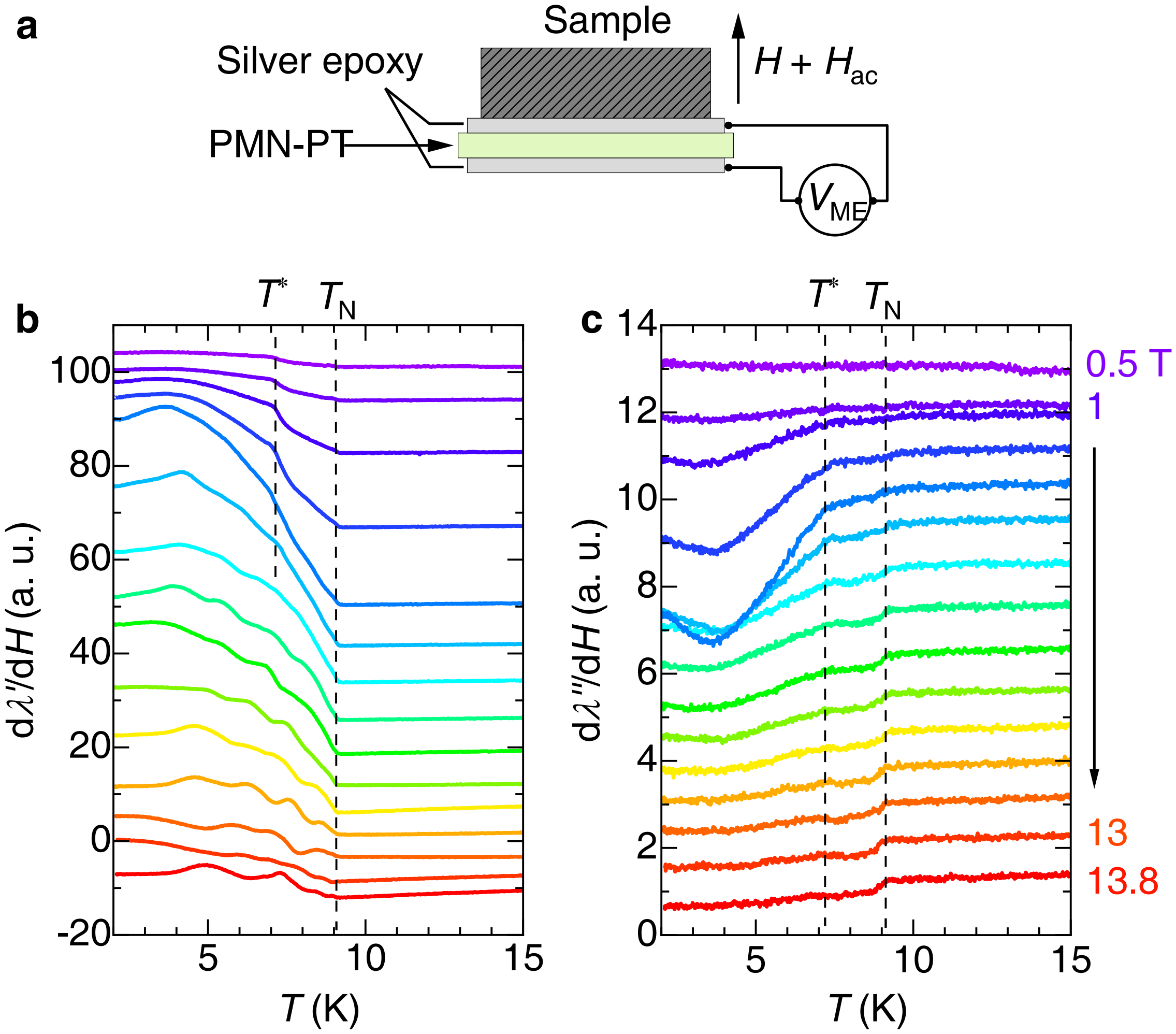

**Fig. 2 | Temperature dependence of the ac magnetostrictive coefficient. a,** Schematic of the composite ME measurement. **b,c,** Temperature dependence of the real (**b**) and imaginary (**c**) components, d$\lambda'$/d$H$ and d$\lambda''$/d$H$, at selected fields. Dashed lines mark the two transition temperatures $T_N$ and $T^*$.

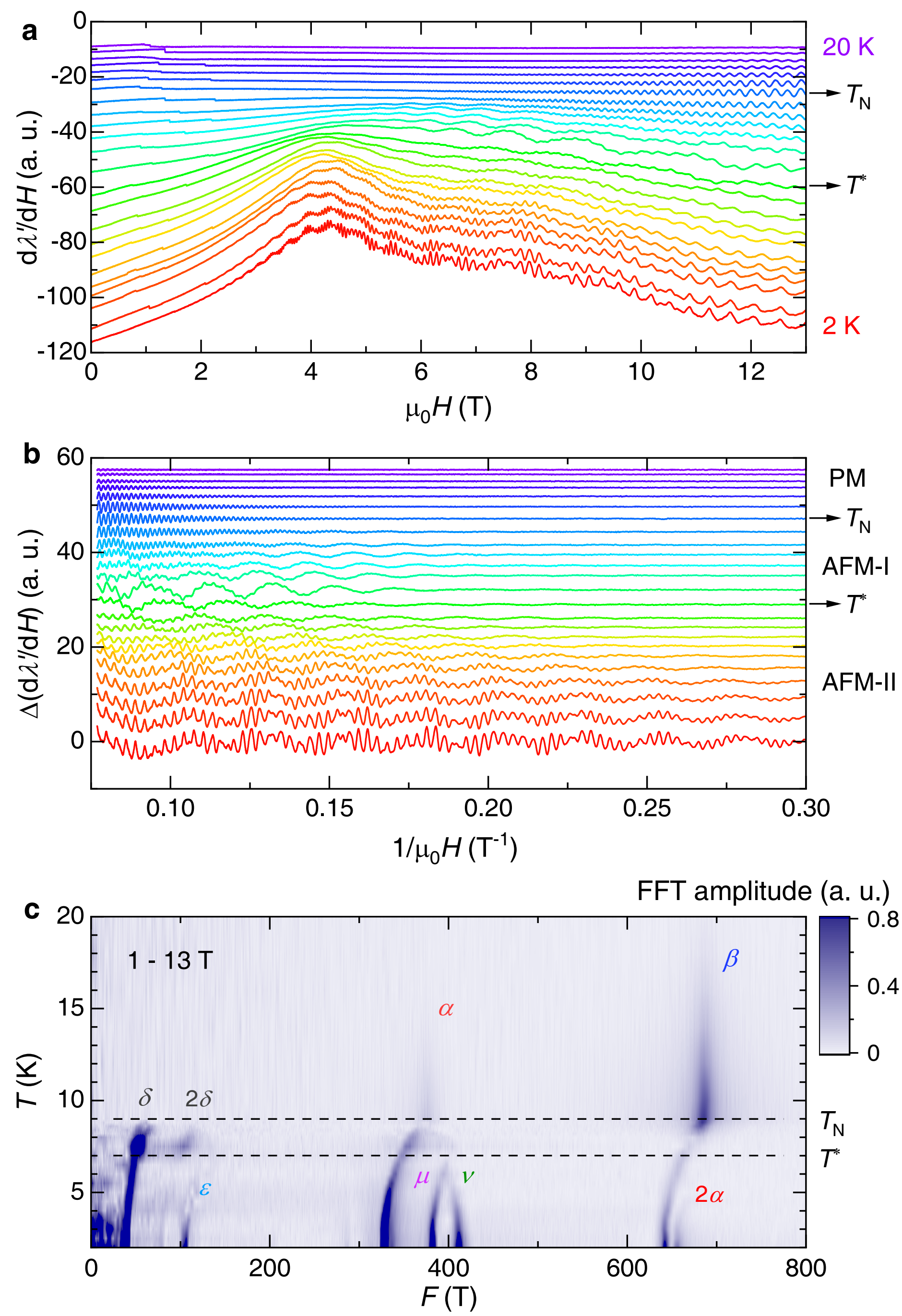

**Fig. 3 | Temperature evolution of the bulk Fermi surface from quantum oscillations. a,** Field-dependent $d\lambda'/dH$ at selected temperatures. **b,** Oscillatory component $\Delta(d\lambda'/dH)$, obtained after subtraction of a smooth magnetic background and plotted against inverse field. **c,** FFT amplitude as a function of frequency and temperature, showing the evolution of the QO branches.

| Phase | **Collinear** | **Type-I** | **Type-II** | **Type-III** |
|---|---|---|---|---|
| MSG | $C2/c.1'_c$ | $Pn\text{-}3m'$ | $R_I\text{-}3$ | $P_I4_132$, $P_I4_332$ |
| MPG | $2/m.1'$ | $m\text{-}3m'$ | $-31'$ | $4321'$ |
| $\mathcal{PT}$ symmetry | Unbroken $\mathcal{PT}$ | Broken $\mathcal{T}$ & $\mathcal{PT}$ | Unbroken $\mathcal{PT}$ | Broken $\mathcal{P}$ & $\mathcal{PT}$ |
| SOC-assisted spin splitting | × | √ | × | √ |

**Fig. 4 | Candidate magnetic structures and symmetry analysis of SmBi.** Collinear, Type-I, Type-II and Type-III AFM configurations are shown together with their magnetic space groups (MSGs) and magnetic point groups (MPGs). The Type-I and Type-III configurations break global $\mathcal{PT}$ symmetry; SOC can therefore lift the residual degeneracy protected by spin-space symmetry.

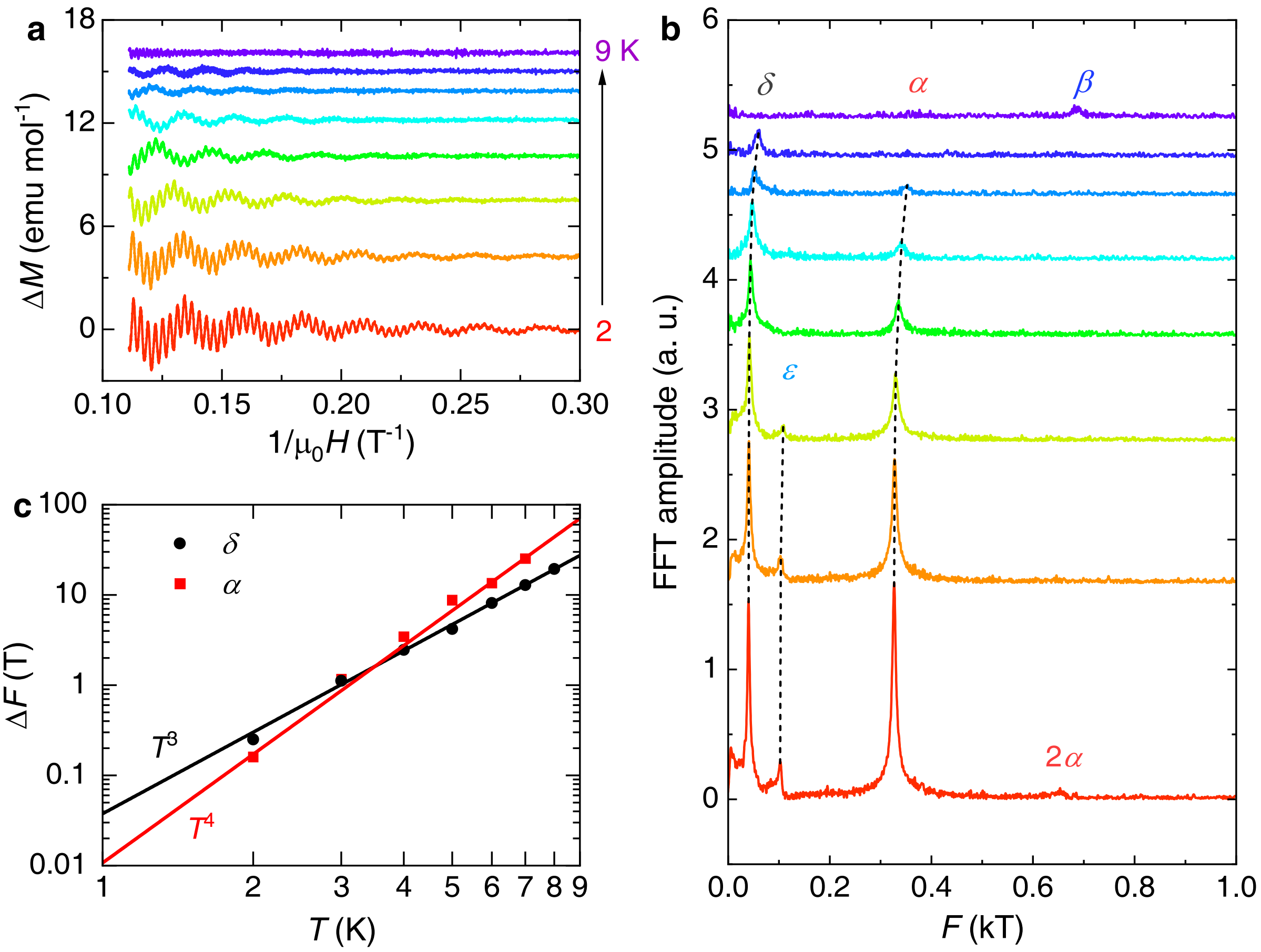

**Extended Data Fig. 1 | de Haas–van Alphen oscillations extracted from the *M*(*H*) curves in Fig. 1c. a,** Oscillatory magnetization plotted against inverse field. **b,** FFT spectra showing four branches, $\delta$, $\varepsilon$, $\alpha$ and $\beta$; $\beta$ is resolved only at 9 K. Dotted guides track the temperature-dependent shifts of $\delta$ and $\alpha$. **c,** Frequency shifts $\Delta F$ of $\delta$ and $\alpha$, which follow $T^3$ and $T^4$ power laws, respectively.

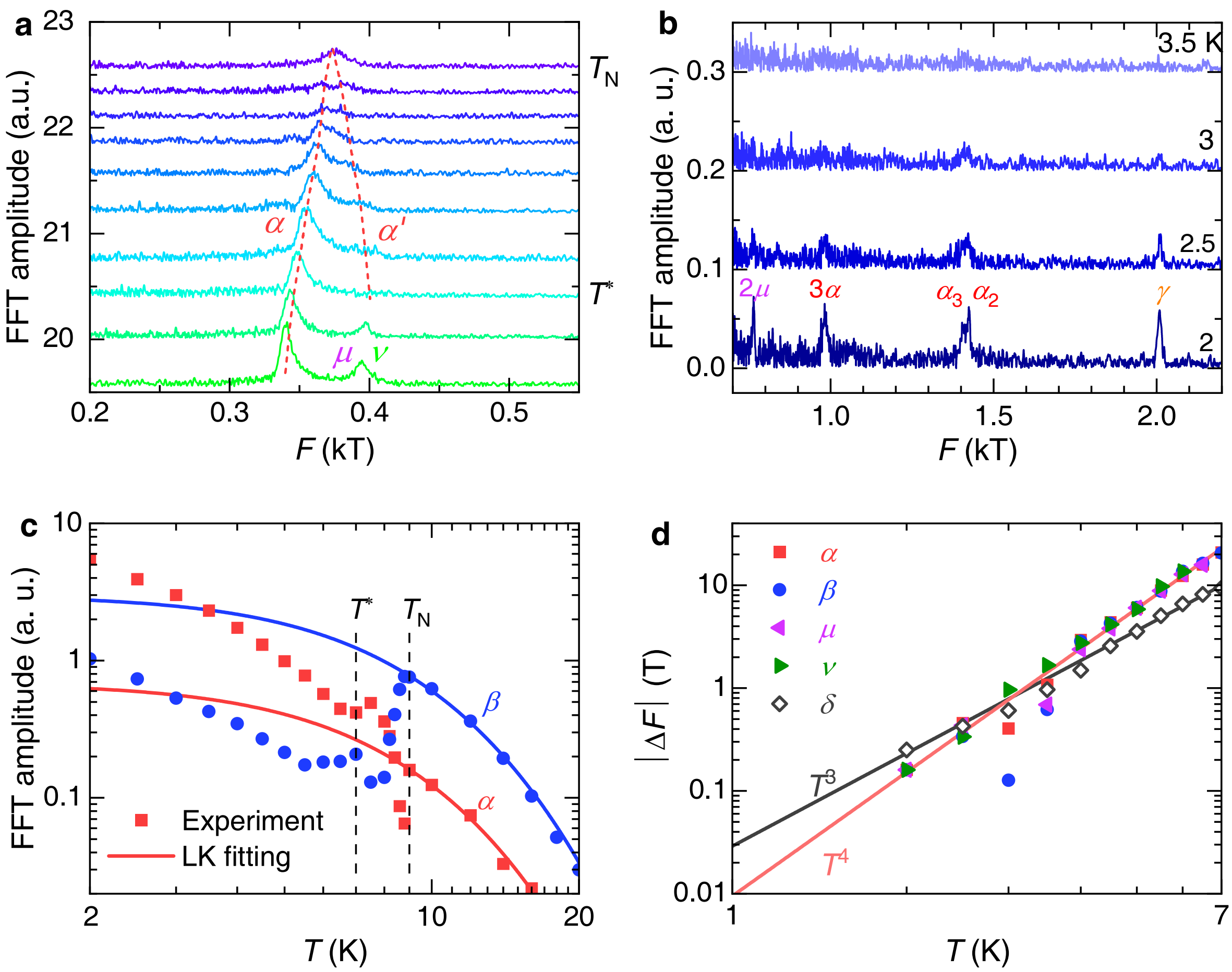

**Extended Data Fig. 2 | Detailed FFT analysis of the QOs in d$\lambda'$/d$H$ from Fig. 3. a,** FFT spectra in the $\alpha$ and $\alpha'$ frequency range. **b,** High-frequency FFT spectra measured between 2 and 3.5 K, revealing features assigned to $\alpha_2$, $\alpha_3$ and $\gamma$. **c,** Temperature-dependent FFT amplitudes of $\alpha$ and $\beta$ from 2 to 20 K. Solid lines are Lifshitz–Kosevich fits to the PM-phase data. **d,** Absolute frequency shift $|\Delta F(T)| = |F(T) - F(0)|$ in AFM-II, where $F(0)$ is the frequency extrapolated to 0 K.

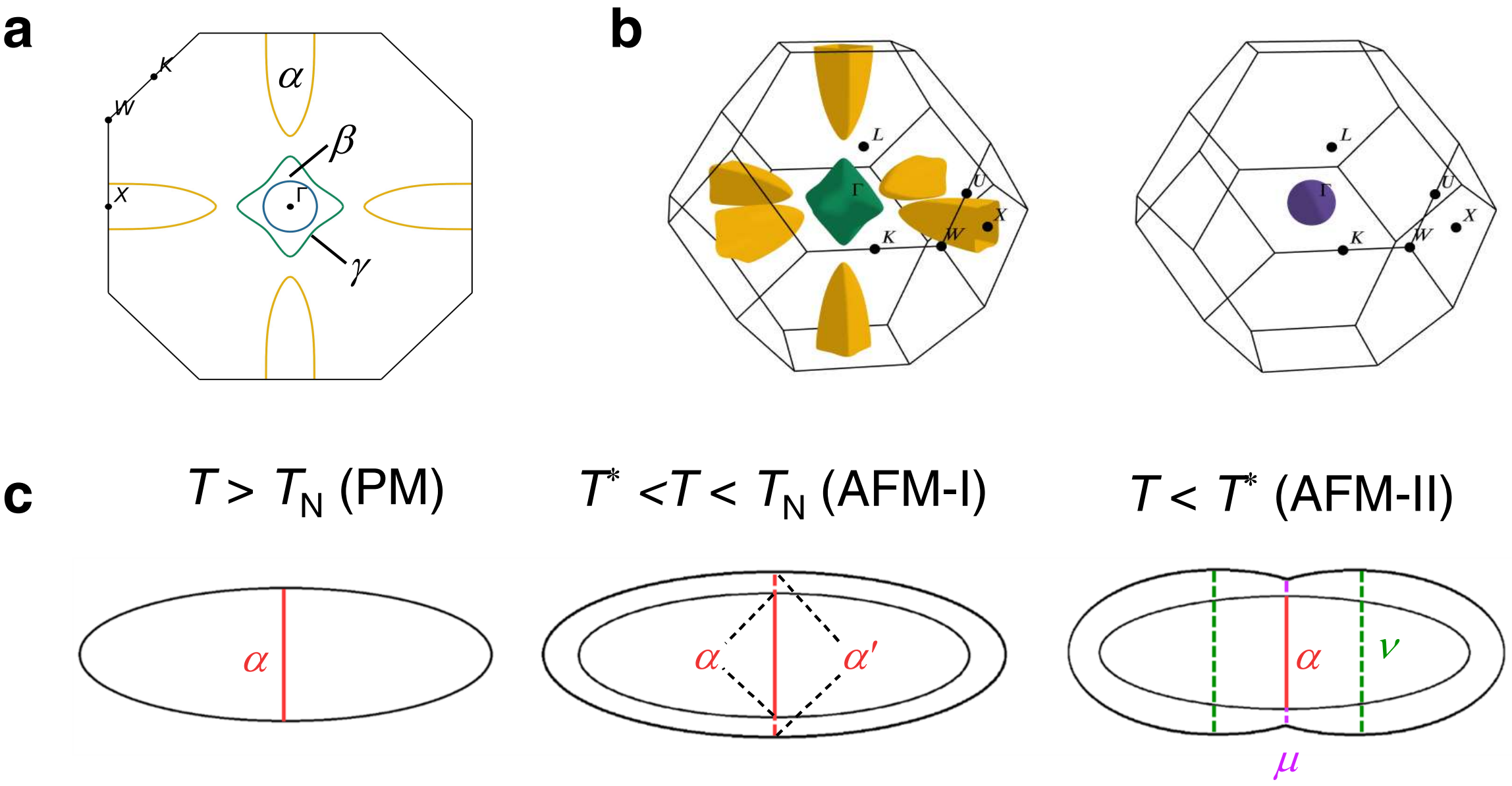

**Extended Data Fig. 3 | Calculated PM Fermi surfaces and proposed evolution of the *α* branch.** **a,** Calculated FS cross-section in the Γ–X–K plane of the Brillouin zone, showing the *α*, *β* and *γ* pockets. **b,** Calculated three-dimensional Fermi surfaces. **c,** Proposed evolution of the *α* electron pocket across the PM, AFM-I and AFM-II phases. The doubly degenerate PM pocket splits into nearby *α* and *α′* ellipsoids in AFM-I through SOC-enabled spin splitting. In the proposed Type-III state of AFM-II, *α′* develops a peanut-like shape with two extremal orbits, *μ* and *ν*.

## Supplementary information

### S1. Proposed magnetic configuration of AFM-I and AFM-II phases.

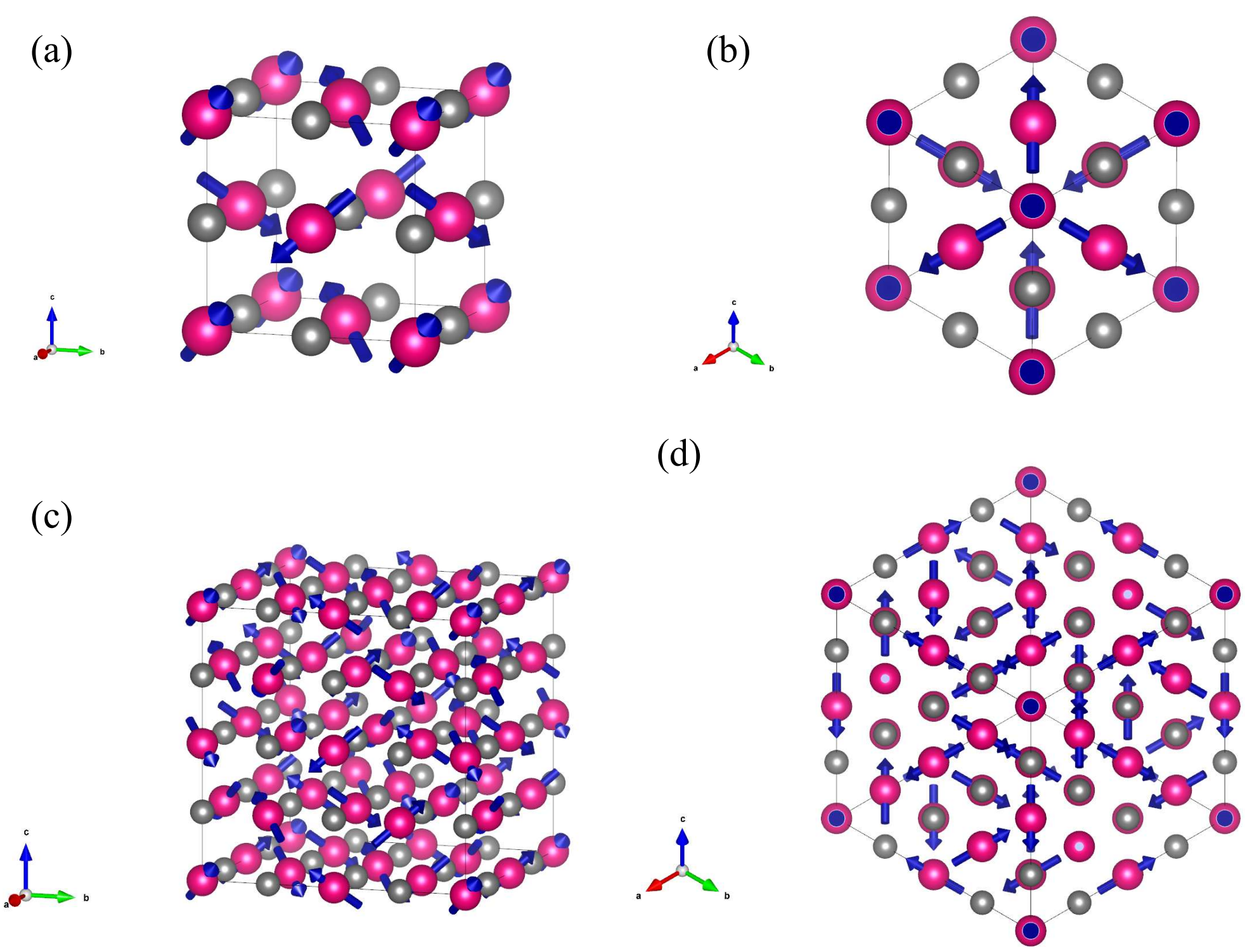

**Figure S1 | Type-I and Type-III magnetic configurations in Ref. 19. a,b,** Type-I proposed for the AFM-I phase; **c,d,** Type-III proposed for the AFM-II phase. The left panels (**a,c**) show the standard crystallographic orientation; the right panels (**b,d**) show the view along [111] direction.

**S2. The dHvA oscillations in SmSb.**

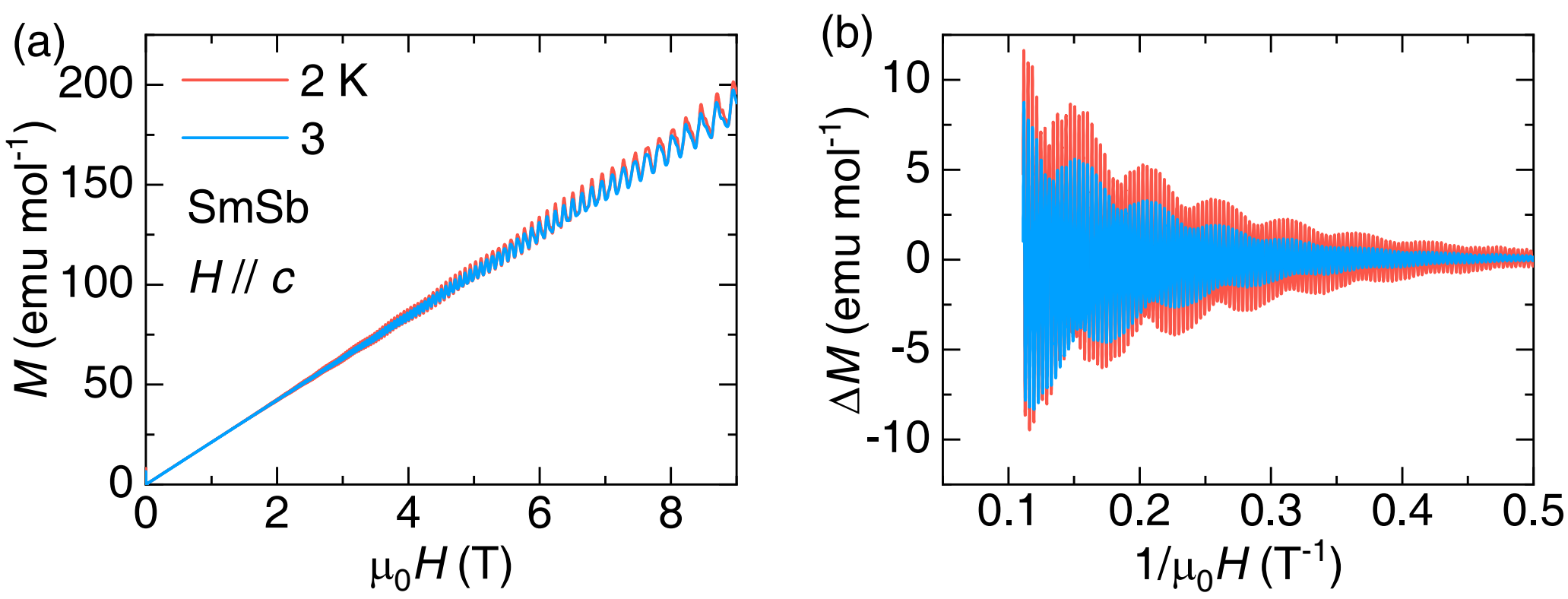

**Figure S2 | de Haas–van Alphen oscillations in SmSb. a,** Field-dependent magnetization $M(H)$ measured at 2 and 3 K with $H$//[001]. **b,** Background-subtracted dHvA oscillations $\Delta M$ plotted as a function of inverse magnetic field $1/\mu_0 H$, revealing clear quantum oscillation patterns.

## S3. SmSb as a $\mathcal{PT}$-preserved benchmark.

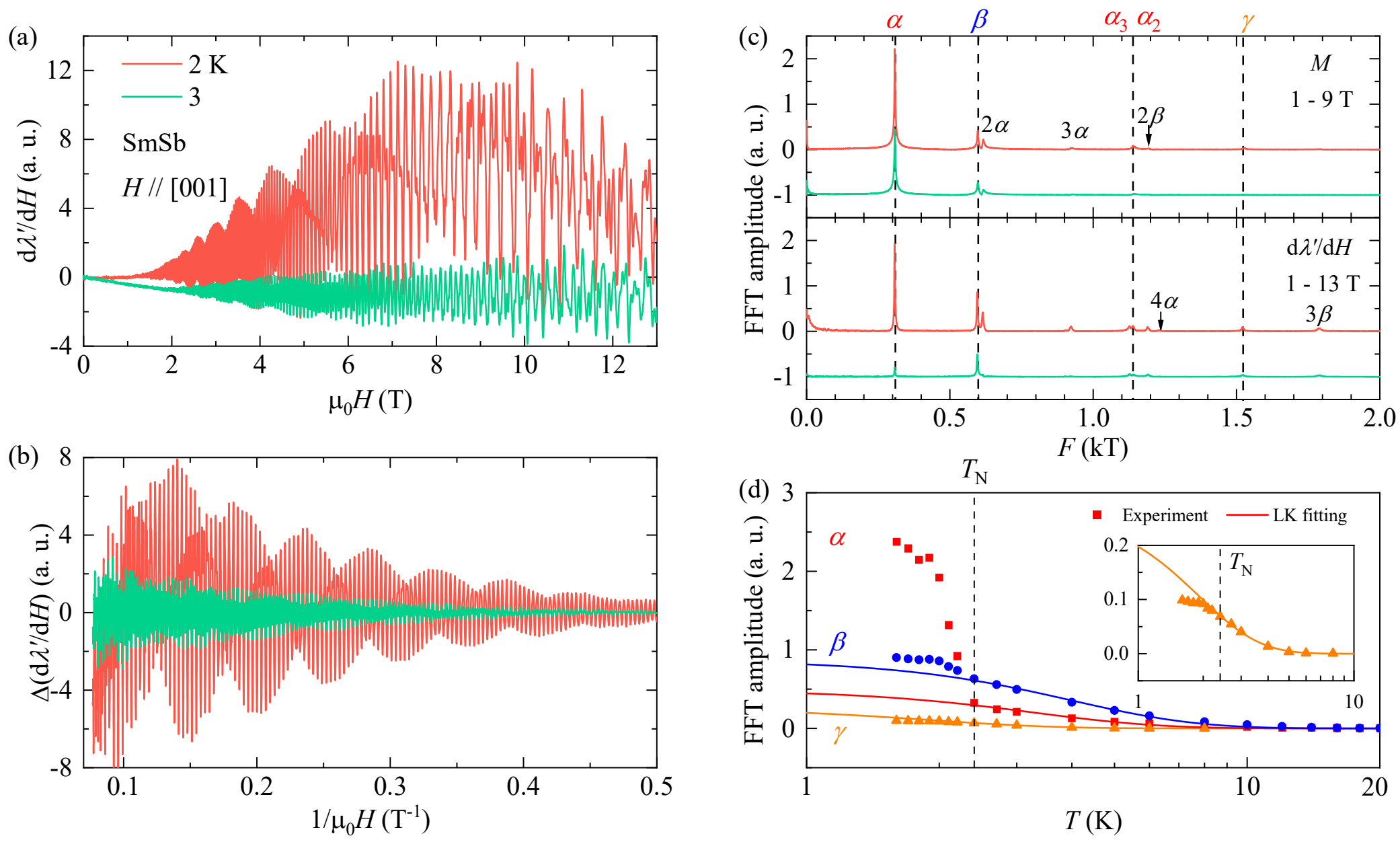

**Figure S3 | SmSb as a $\mathcal{PT}$-preserved benchmark. a,** The field-dependent d$\lambda'$/d$H$ in SmSb at 2 K (AFM) and 3 K (PM) with $H$//[001]. **b,** Quantum oscillations extracted by subtracting smooth backgrounds in the inverse magnetic field. **c,** Top panel: FFT spectra of dHvA oscillations in Fig. S2 with FFT windows of 1-9 T, Bottom panel: FFT spectra of the oscillations in (**b**) with FFT windows of 1-13 T. Harmonic oscillation frequencies are labeled in black. **d,** The $T$-dependent FFT amplitude of the d$\lambda'$/d$H$ oscillations with the selected temperatures from 1.6 to 20 K. For $T > T_N$, $\alpha$ (red), $\beta$ (blue), and $\gamma$ (orange) are fit with LK behaviors (solid lines). For $T < T_N$, all of them deviate from the LK behaviors. The inset panel shows the divergence behavior for $\gamma$.

We establish SmSb as the "benchmark" case in which the ordered state leaves the bulk Fermi surface essentially unchanged. Single crystals SmBi were grown by a self-flux method. Figure S3(a) displays d$\lambda'$/d$H$ of SmSb versus $H$ for representative temperatures (2 and 3 K are shown as representative data for AFM and PM phase, respectively). Clear quantum oscillations set in at fields as low as 1.3 T at 2 K—well below typical onset fields in SdH oscillations (Ref. 11). In both AFM and PM states, the non-oscillatory background in d$\lambda'$/d$H$ remains weak, much smaller than that in resistivity (Ref. 11), yielding clean oscillation patterns that persist up to 20 K.

The background-subtracted magnetostrictive coefficients $\Delta$(d$\lambda'$/d$H$) plotted against 1/$B$ [Fig. S3(b)] reveals multiple frequencies. FFT spectra [bottom panel of Fig. S3(c)] resolve at least eleven sharp peaks at 2 K. Orbits $\alpha$ (308 T), $\beta$ (596 T), and $\gamma$ (1523 T) are fundamental ones, while the remaining peaks are assigned to harmonics up to fourth order, consistent with Ref. 11 and our dHvA measurement [top panel of Fig. S3(c)]. Note that, orbits $\alpha_2$, $\alpha_3$, and $\alpha$ stem from the identical but mutually orthogonal electron ellipsoids at the Brillouin zone boundary. A slight tilt of field away from the [001] direction breaks the degeneracy of $\alpha_2$ and $\alpha_3$ orbits, as reported previously (Ref. 12). Within the ultrahigh resolution of the composite ME method, no additional fundamental peaks are resolved. At 3 K, the same three fundamental frequencies are observed with no discernible shift, indicating that the Fermi surfaces of SmSb are essentially unchanged across the AFM–PM boundary. This is consistent with a collinear AFM order where $\mathcal{PT}$ symmetry is preserved, protecting the Kramers-like degeneracy of the bands.

In addition, the oscillation amplitude changes markedly across $T_N$. Fig. S3(d) summarizes the temperature dependence of the FFT amplitudes from 1.6 to 20 K. Above $T_N$, all the fundamental orbits follow the Lifshitz–Kosevich (LK) behavior, yielding effective masses $m^*$=0.097(7) , 0.080(3), and 0.166(5) $m_0$ ($m_0$ is the free-electron mass) for $\alpha$, $\beta$, and $\gamma$, respectively. Below $T_N$, QO amplitudes of $\alpha$ and $\beta$ are strongly enhanced and deviate from LK expectations, while that of $\gamma$ shows a weaker anomaly near $T_N$ and falls slightly below the LK extrapolation from the PM side, indicative of critical scattering or mass renormalization. Similar deviations have been noted in SdH but not in dHvA oscillations of SmSb (Ref. 11 & 12). Because the ME technique is a bulk probe of the magnetostrictive coefficient, surface contributions can be ruled out as the origin of these non-LK behaviors in SmSb.

**S4. Effective mass fitting for SmBi.**

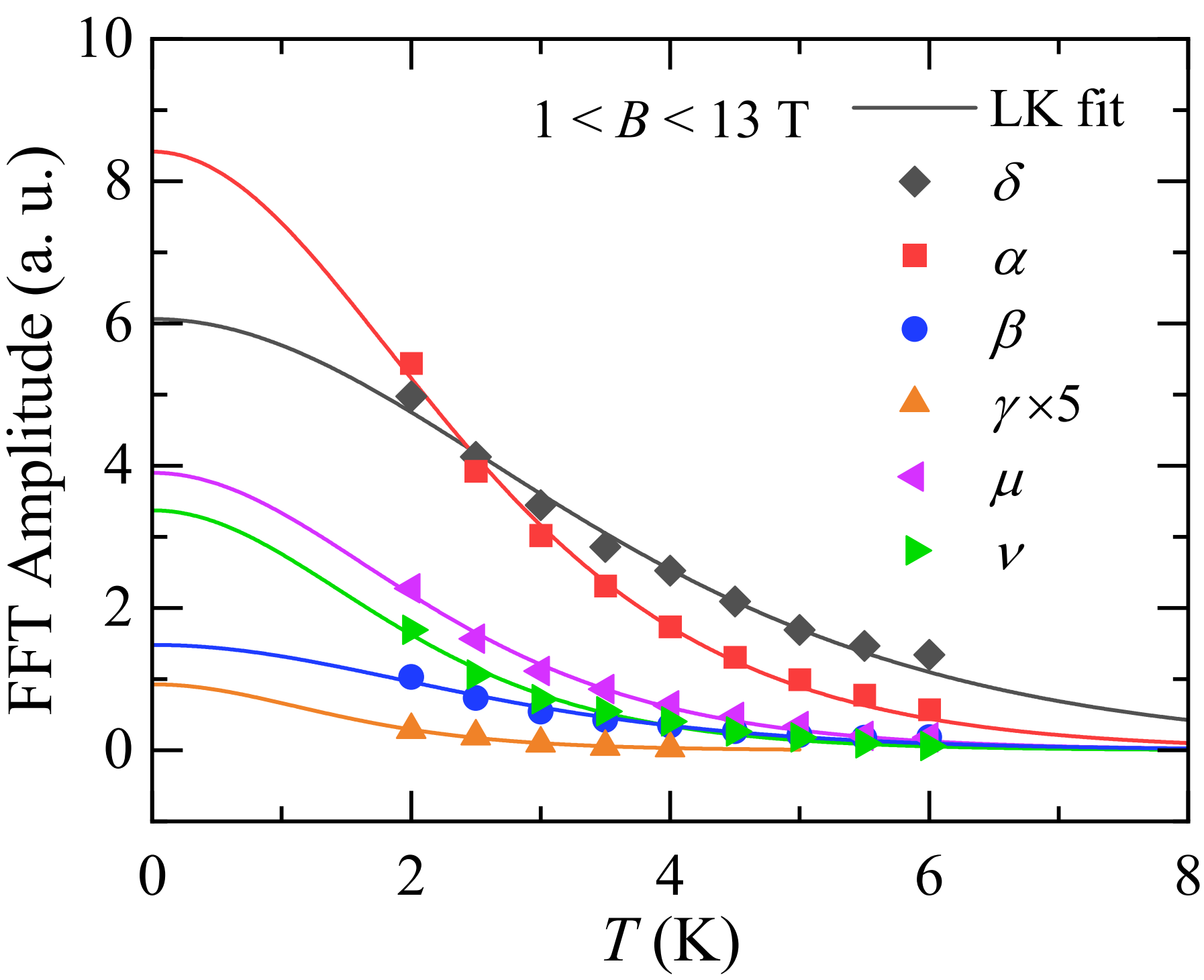

**Figure S4 | Effective mass fitting for SmBi.** Temperature-dependent FFT amplitudes for the $\delta$, $\alpha$, $\beta$, $\gamma$, $\mu$ and $\nu$ orbits, respectively. Solid lines represent fits to thermal damping factor in the Lifshitz–Kosevich formula, yielding effective masses $m^*$ = 0.078(3), 0.112(3), 0.106(6), 0.185(10), 0.124(4) and 0.142(6) $m_0$, respectively, where $m_0$ is the free-electron mass.

**S5. Band calculation of SmBi with the magnetic moment of Sm 4*f* electron considered**

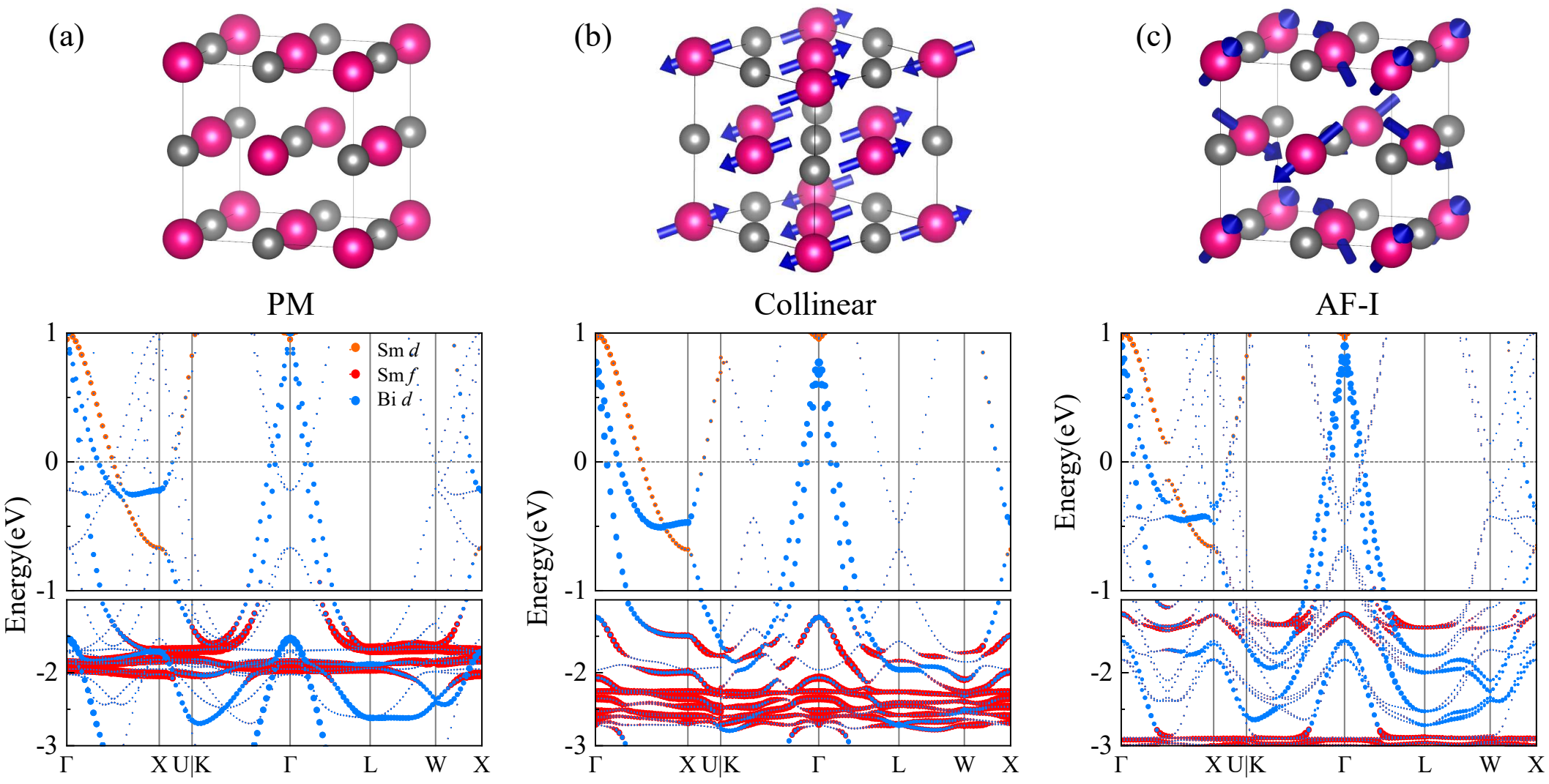

**Fig. S5 | Band structure calculations for three magnetic configurations of SmBi.** To elucidate the effect of magnetic order on the electronic structure of SmBi, we performed density functional theory (DFT) calculations including Sm 4*f* electrons with fixed magnetic moments. Three magnetic configurations were examined: **a,** Paramagnetic (PM) state. **b,** Collinear antiferromagnetic (AFM) state. **c,** Noncollinear Type-I state. The top panels illustrate the magnetic spin configurations on the Sm sublattice. The bottom panels show the corresponding DFT-calculated band structures along high-symmetry paths in the Brillouin zone, with all bands unfolded to the primitive cell. All calculated band structures were unfolded back to the primitive-cell Brillouin zone for direct comparison.

**Fig. S5(a) – PM state.** In the PM configuration, the system preserves both spatial inversion ($\mathcal{P}$) and time-reversal ($\mathcal{T}$) symmetries. The band structure exhibits complete spin degeneracy throughout the Brillouin zone, as shown in the calculated bands. **Fig. S5(b) – Collinear AFM state.** In this configuration, the magnetic order breaks certain symmetries but preserves the combined $\mathcal{PT}$ symmetry. The magnetic structure belongs to the magnetic space group $C2/c.1'_c$ and magnetic point group $2/m.1'$. As a consequence, the band structure remains fully spin-degenerate, similar to the PM case. No additional band splitting or Fermi surface reconstruction is observed. **Fig. S5(c) – Noncollinear Type-I state (Kouvel–Kasper type).** This configuration forms without altering the chemical unit cell (remaining cubic). Its magnetic structure is classified as magnetic space group *Pn*-3*m*′ and magnetic point group *m*-3*m*′. The $\mathcal{T}$ symmetry is bonded to a mirror symmetry, resulting in broken $\mathcal{PT}$ symmetry. In stark contrast to the PM and collinear cases, the calculated band structure reveals global spin splitting across the Brillouin zone, particularly around the Γ point. This splitting originates from the synergistic effect of the noncollinear magnetic order and strong SOC, and provides the theoretical basis for the SOC-enabled Fermi surface reconstruction observed experimentally in the AFM-I phase of SmBi.

**Detailed discussion of calculation methods**

The electronic structure of Sm compounds presents a significant challenge due to the dual nature of 4*f* electrons. In our computational framework, we adopt a two-tier strategy to handle the Sm valence states depending on the magnetic configuration under study.

A. Valence Configuration and POTCAR Selection.

For the non-magnetic (NM) reference state, the 4*f* electrons are considered to be deep-core states that do not participate in the itinerant bonding near the Fermi level. Using a Sm potential with 11 valence electrons effectively stabilizes the convergence by avoiding the numerical instabilities associated with highly localized and strongly correlated 4*f* orbitals in a non-polarized manifold. However, in the AFM phase, the 4*f* electrons are the primary source of the magnetic moments and the driving force behind the observed altermagnetic spin-splitting. Therefore, the explicit inclusion of 4*f* states (16 valence electrons) is mandatory for AFM calculations to capture the exchange interactions and the resulting *k*-space spin textures.

B. Calibration of the Fermi Level.

Experimental comparisons revealed a discrepancy in the Fermi surface topology when shifting from 11-electron to 16-electron potentials. The introduction of the 4*f* manifold in the 16-electron calculation often leads to a self-consistent shift of the Fermi level ($E_F$) that does not align with the itinerant band filling observed in experimental transport or NM reference results. This is primarily due to the complex charge re-distribution between the localized 4*f* and itinerant *s*-*d* bands.

Thus, we implemented a corrective alignment procedure. We used the electron distribution of the NM 11-electron calculation as a rigid reference for the itinerant band filling. The Fermi level of the AFM band structure was then adjusted to match this reference filling. This ensures that the energy bands of interest are compared at the same chemical potential, allowing for an accurate evaluation of the symmetry-protected degeneracies and the SOC-induced splitting without artifacts arising from 4*f*-related occupancy shifts.

**S6. Distinct power law in temperature-dependent oscillatory frequency in SmBi**

One striking feature in SmBi is the power laws hidden in the temperature dependence of the oscillation frequencies in the AFM-II phase, as shown in Extended Data Fig. 2d. On one hand, $\alpha$, $\beta$, $\mu$, and $\nu$ exhibit nearly identical $\Delta|F|$ scaling, well described by a $T^4$ law with coefficient ~9.5 × $10^{-3}$ T/$K^4$. On the other hand, the $\delta$ branch follows a distinct $T^3$ law with coefficient ~2.9 × $10^{-2}$ T/$K^3$. While $T^4$ frequency shifts have been reported in a few nonmagnetic metals and attributed to thermal expansion and electron–phonon coupling. Furthermore, the expected magnitudes, e.g., ∼3×$10^{-6}$ T/$K^4$ in Au, are orders of magnitude smaller than observed here, ruling out such nonmagnetic origins for SmBi. A $T^3$ dependence is even rarer. The distinct power laws likely reflect different band topologies (e.g., linear vs. parabolic bands) coupling to order parameter of the AFM-II phase.

**S7. Summary of quantum oscillation frequencies and effective masses for SmSb and SmBi in the PM and AFM states.**

| | orbit | $F_{PM}$(T)[1] | $m^*(m_0)$ | $F_{AFM}$(T)[2] | $m^*(m_0)$ |
|---|---|---|---|---|---|
| SmSb | $\alpha$ | 308 | 0.097(7) | 308 | |
| | $\beta$ | 596 | 0.080(3) | 596 | |
| | $\gamma$ | 1523 | 0.166(5) | 1522 | |
| | $\alpha_3$ | 1128 | | 1127 | |
| | $\alpha_2$ | 1140 | | 1140 | |
| SmBi | $\alpha$ | 376 | 0.047(2) | 328 | 0.112(3) |
| | $\beta$ | 686 | 0.045(1) | 642 | 0.106(6) |
| | $\gamma$ | | | 2009 | 0.185(10) |
| | $\alpha_3$ | | | 1403 | |
| | $\alpha_2$ | | | 1424 | |
| | $\delta$ | | | 40 | 0.078(3) |
| | $\varepsilon$ | | | 106 | |
| | $\mu$ | | | 382 | 0.124(4) |
| | $\nu$ | | | 412 | 0.142(6) |

[1]PM state data: measured at $T = 3$ K for SmSb and $T = 12$ K for SmBi.

[2]AFM state data: measured at $T = 2$ K for both compounds. SmBi is in the AFM-II phase.